\documentclass[conference,10pt]{IEEEtran}
\usepackage{epsfig,makeidx,color,epstopdf}
\usepackage{amsmath,amssymb,amsthm,bbm}
\usepackage{cite,graphicx}
\usepackage{enumerate}

\def\rH{{\rm H}}
\def\rT{{\rm T}}

\def\uE{{\mathbb E}}

\def\be{ \begin{equation} }
\def\ee{ \end{equation} }
\def\bea{ \begin{eqnarray} }
\def\eea{ \end{eqnarray} }
\def\bx{{\bf x}}

\def\ba{{\bf a}}

\def\bh{{\bf h}}
\def\bp{{\bf p}}

\def\bw{{\bf w}}

\def\bR{{\bf R}}

\def\b0{{\bf 0}}

\def\cC{{\cal C}}

\def\cN{{\cal N}}

\def\sI{{\sf I}}

\def\sI{{\sf I}}

\ifCLASSOPTIONonecolumn
  \newcommand{\figwidth}{0.55\columnwidth}
\else
  \newcommand{\figwidth}{0.90\columnwidth}
\fi

\IEEEoverridecommandlockouts

\begin{document}

\title{NOMA: Principles and Recent Results}

\author{Jinho Choi\\
School of EECS, GIST \\
\emph{Email: jchoi0114@gist.ac.kr} \\
(Invited Paper)
\thanks{This work was supported by
the ``Climate Technology Development and Application"
research project (K07732) through a grant provided by GIST in 2017.
}}

\date{today}
\maketitle

\begin{abstract}
Although 
non-orthogonal multiple access (NOMA) 
is recently considered for cellular systems,
its key ideas such as successive
interference cancellation (SIC) and superposition coding
have been well studied in information theory.
In this paper, we overview principles of 
NOMA based on information theory and present some recent results.
Under a single-cell environment, 
we mainly focus on fundamental issues,
e.g., power allocation and beamforming
for downlink NOMA and coordinated and uncoordinated
transmissions for uplink NOMA.
\end{abstract}


\section{Introduction}

Recently, nonorthogonal multiple access (NOMA)
has been extensively studied in \cite{Saito13}
\cite{Kim13} \cite{Perotti14} \cite{Ding14}
for 5th generation (5G) systems as NOMA can improve
the spectral efficiency.
In order to implement NOMA within standards,
multiuser superposition transmission
(MUST) schemes are proposed in  \cite{3GPP_MUST}. 
There are also review articles for NOMA, e.g., \cite{Dai15} \cite{Ding_CM}.

Although the application of
NOMA to cellular systems is relatively new,
NOMA is based on well-known schemes such as superposition
coding and successive interference cancellation (SIC) 
\cite{Wyner74}  \cite{CoverBook}.
In particular, the decoding 
approach based on multiple single-user
decoding with SIC for multiple access channels 
has been studied from the information-theoretic point of view
under other names such as stripping and onion peeling \cite{Gallager94}.
In addition, there are precursors of NOMA. For example,
code division multiple access (CDMA) is a NOMA scheme
as spreading codes are not orthogonal \cite{Kohno95} \cite{ViterbiBook}
\cite{Adachi98}. 
In order to mitigate
the multiple access interference (MAI) due to non-orthogonal
spreading codes, multiuser detection (MUD) is also studied
\cite{VerduBook} \cite{ChoiJBook2}.
In CDMA, although the notion of superposition coding
is not actively exploited, SIC 
is extensively studied since \cite{Patel94}.

The main difference of the NOMA schemes for 5G from existing 
CDMA schemes is the exploitation of the power difference
of users and the asymmetric application of SIC in the power
and rate allocation.
In particular, these features are well shown in downlink NOMA.
In \cite{Saito13} \cite{Kim13}, a user close to a
base station (BS) and a user far away from the BS
form a group or cluster. For convenience,
the former and latter users are called strong and weak users,
respectively (in terms of their channel gains).
It is expected to transmit a higher power to the weak user
than the strong user due to the path loss or channel gain.
If they share the same radio resource block,
the signal to the weak user received at the strong user
has a higher signal-to-noise ratio (SNR)
than that at the weak user,
which implies that the strong user is able to decode
the signal to the weak user and remove it (using SIC) to decode 
the desired signal without MA without MAI.
On the other hand, at the weak user, the signal to the strong
user is negligible as its transmission power is lower
than that to the weak user. Thus, the weak user decodes
the desired signal without using SIC.

To exploit the power difference, the power allocation becomes crucial.
A power allocation problem
for NOMA with fairness is studied in \cite{Timotheou15} \cite{Choi16a}. 
An energy efficient
power allocation approach is investigated in \cite{Fang16}.
The power difference between users 
can also be exploited in a multi-cell system.
In \cite{Choi14}, NOMA is studied for downlink coordinated
two-point systems. In \cite{Shin17},
coordinated beamforming is considered for multi-cell NOMA.

In \cite{Sun15} \cite{Ding16} \cite{Choi16_5}, 
multiple input multiple output (MIMO)
for NOMA is studied to see how NOMA can be applied to MIMO systems.
Beamforming in NOMA is also studied in
\cite{Kim13} \cite{Higuchi13} \cite{Higuchi15}.
In general, beamforming in NOMA is to exploit the power and 
spatial domains. 
In \cite{Ding16}, beamforming with limited feedback
of channel state information (CSI)
is studied.
Multicast NOMA beamforming is considered in \cite{Choi15}.


Since there have been various NOMA schemes and related approaches,
it might be important to have an overview. 
As mentioned earlier, there are already excellent review
articles \cite{Dai15} \cite{Ding_CM}.
However, they focus on system aspects. Thus,
in this paper, we aim at providing an overview of key approaches
and recent results with emphasis on
fundamentals of NOMA under a single-cell environment.

The rest of the paper is organized as follows.
In Section~\ref{S:SM}, we present system models for uplink
and downlink NOMA.
The power allocation problem
and downlink beamforming are considered for 
downlink NOMA in Section~\ref{S:PA}
and Section~\ref{S:DB}, respectively.
We focus on some key issues in uplink NOMA in Section~\ref{S:UN}
and conclude the paper with some remarks in Section~\ref{S:Conc}.

\subsubsection*{Notation}
Matrices and vectors are denoted by upper- and lower-case
boldface letters, respectively.
The superscripts $*$, $\rT$, and $\rH$
denote the complex conjugate, transpose, Hermitian transpose, respectively.
$\uE[\cdot]$ and ${\rm Var}(\cdot)$
denote the statistical expectation and variance, respectively.
$\cC \cN(\ba, \bR)$
represents the distribution of
circularly symmetric complex Gaussian (CSCG)
random vectors with mean vector $\ba$ and
covariance matrix $\bR$.

\section{System Models}	\label{S:SM}

\subsection{Downlink NOMA}

In this section, we present a system model 
consisting of a BS and multiple users for downlink NOMA.
Throughout the paper, we assume that the BS and users are equipped
with single antennas.

Suppose that there are $K$ users in the same resource block for downlink.
Let $s_{k,t}$ and $h_k$ denote the data symbol at time $t$ and
channel coefficient from the BS to user $k$, respectively.
The block of data symbols, $[s_{k,0} \ \ldots \ s_{k,T-1}]^\rT$,
where $T$ is the length of data block,
is assumed to be a codeword of a capacity-achieving code.
Furthermore, $T$ is shorter than the coherence time
so that $h_k$ remains unchanged for the duration of a block transmission.
Suppose that 
superposition coding \cite{CoverBook} is employed for NOMA and the signal
to be transmitted by the BS is $\sum_{k=1}^K s_{k,t}$. 
Then, at user $k$, the received signal is given by
\be
y_{k,t} = h_k \sum_{m=1}^K s_{m,t} + n_{k,t},
t = 0,\ldots, T-1,
	\label{EQ:yk}
\ee
where $n_{k,t} \sim \cC \cN(0, 1)$ is the independent
background noise
(here, the variance of $n_{k,t}$ is normalized for convenience).
Let $\alpha_k = |h_k|^2$ and $P_k = \uE[|s_{k,t}|^2]$ 
(with $\uE[s_{k,t}] = 0$). Then,
$P_k$ becomes the transmission power allocated to $s_{k,t}$.
In this section, we assume that the BS knows all
the channel gains, $\{\alpha_k\}$, and studies the power allocation
to enable SIC at users for NOMA.

\subsection{Uplink NOMA}	\label{S:UN}

We again assume that there are $K$ users who are allocated
to the same resource block for uplink transmissions.
Then, at the BS, the received signal becomes
\be
r_{t} = \sum_{k=1}^K g_k u_{k,t} + n_t, \ t = 0, \ldots, T-1,
	\label{EQ:rt}
\ee
where $g_k$ and $u_{k,t}$ 
represent the channel coefficient from the $k$th
user to the BS and the signal from the $k$th user,
respectively, and $n_t \sim \cC \cN(0, 1)$ is the background
noise at the BS.
The BS is able to decode all $K$ signals if 
the transmission rates and powers of the $u_{k,t}$'s are properly
decided using single-user decoding.
A well-known example of uplink NOMA
is CDMA where each user's signal
is spread signal with a unique spreading code \cite{ViterbiBook}.

\section{Power Allocation for Downlink NOMA}	\label{S:PA}

In this section, we briefly study the power allocation for downlink 
NOMA with achievable rates.

Let $R_k$ denote the transmission rate of $s_{k,t}$
in \eqref{EQ:yk}
and $C_{l;k}$ denote the achievable rate 
for the signal to user $l$ at user $k$ with SIC
in descending order, where $l \ge k$.
Then, it can be shown that
\be
C_{l;k} = \log_2 \left(
1 + \frac{\alpha_k P_l}{ \alpha_k \sum_{m=1}^{l-1} P_m + 1}
\right).
\ee
Assume that user $k$ has to decode his/her signal
(i.e., $\{s_{k,t}\}$)
as well as the signals to user 
$l$, $l \in \{k, \ldots, K\}$ for SIC in NOMA. 
In this case, the rate-region can be found as
\be
R_l < C_{l;k}, \ l \in \{k, \ldots, K\}, \ k \in \{1,\ldots, K\}.
	\label{EQ:RC}
\ee
Clearly, user $k$ should be able to decode the signals 
to users $k, \ldots, K$.
As an example, suppose that $K = 2$.
At user 1, the signal to user 2 can be decoded if
$$
R_2 < C_{2;1} = \log_2 \left( 1+ \frac{\alpha_1 P_2}{\alpha_1 P_1 + 1}
\right).
$$
Once $\{s_{2,t}\}$ is decoded at user 1 and removed
by SIC, 
$C_{1;1}$ becomes $\log_2 \left( 1+ \alpha_1 P_1 \right)$
and the signal to user 1 can be decoded if
\be
R_1 < C_{1;1} = \log_2 \left( 1+ \alpha_1 P_1 \right).
	\label{EQ:RC1}
\ee
At user 2, assuming that $\{s_{1,t}\}$ is sufficiently weaker than 
$\{s_{2,t}\}$, $\{s_{2,t}\}$ can be decoded if
\be
R_2 < C_{2;2} = 
\log_2 \left( 1+ \frac{\alpha_2 P_2}{\alpha_2 P_1 + 1}
\right).
	\label{EQ:RC2}
\ee
If $\alpha_1 \ge \alpha_2$ is assumed, we have
$$
\frac{\alpha_2 P_2}{\alpha_2 P_1 + 1}
\le \frac{\alpha_1 P_2}{\alpha_1 P_1 + 1}
\ \mbox{or} \ C_{2;2} \le C_{2;1}.
$$
Thus, the rate-region of $R_1$ and $R_2$ in \eqref{EQ:RC} is reduced
to 
\begin{align*}
R_1 < C_{1;1} \ \mbox{and} \ R_2 < C_{2;2} .
\end{align*}
In general, for any $K \ge 2$, if 
\be
\alpha_1 \ge \ldots \ge \alpha_K,
	\label{EQ:alps}
\ee
the rate-region in \eqref{EQ:RC} is reduced to 
\begin{align}
R_l & < C_{l;l}, \ l = 1, \ldots, K,
	\label{EQ:PAP}
\end{align}
because $C_{l;k} \ge C_{l;l}$, $k \in \{1,\ldots,l-1\}$,
$l \in \{1,\ldots,K\}$.

If the BS knows 
CSI and orders the users according to \eqref{EQ:alps},
we can consider a power allocation problem for downlink NOMA to 
maximize the sum rate subject to a total power constraint as follows:
\begin{eqnarray}
& \max_\bp \sum_{k=1}^K C_{k;k} & \cr
& \mbox{subject to} \ \sum_{k=1}^K P_k \le P_{\rm T}, & 
	\label{EQ:PA2}
\end{eqnarray}
where $\bp = [P_1 \ \ldots \ P_K]^\rT$ and
$P_{\rm T}$ is the total transmission power.
The power allocation problem in \eqref{EQ:PA2} 
has a trivial solution that is
$P_1 = P_{\rm T}$ and $P_k = 0$, $k = 2,\ldots, K$.
To avoid this, rate constraints can be taken into account.
To this end,
we can consider another power allocation problem
to minimize the total transmission power 
with rate constraints as follows:
\begin{eqnarray}
& \min_\bp ||\bp ||_1  & \cr
& \mbox{subject to} \ C_{k;k} \ge \bar R_k, \ k = 1,\ldots,K, & 
	\label{EQ:PA1}
\end{eqnarray}
where $\bar R_k$ represents the (required) minimum rate for user $k$
and $||\bx||_1 = \sum_i |x_i|$
denotes the 1-norm of vector $\bx$.
This problem formulation 
can be employed instead of the approaches in
\cite{Choi16_F} and \cite{Timotheou15} for fairness as 
each user can ensure a guaranteed rate, $\bar R_k$.
In Appendix~\ref{A:1}, we provide the solution to \eqref{EQ:PA1}.

\begin{figure}[thb]
\begin{center}
\includegraphics[width=\figwidth]{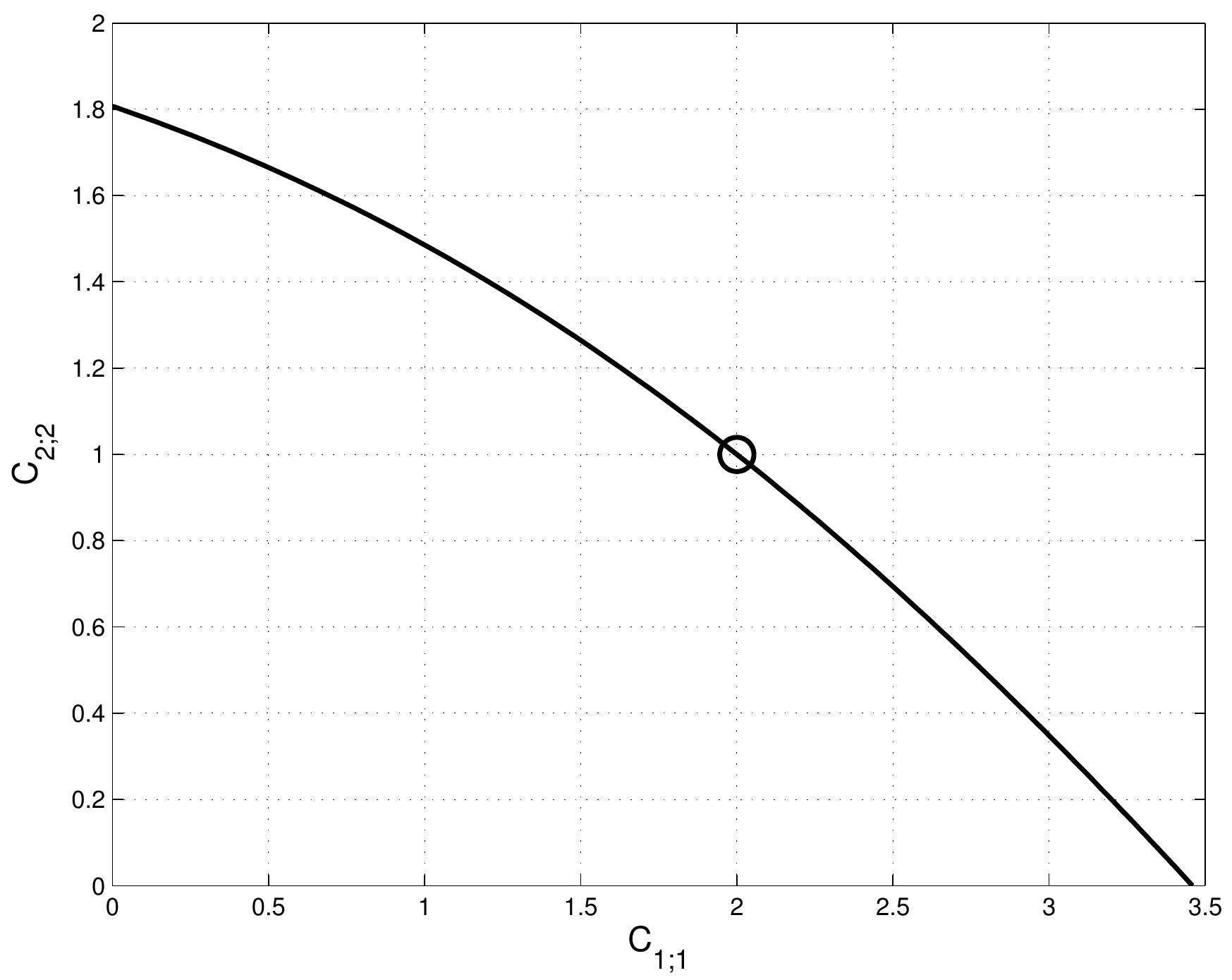} \\
\end{center}
\caption{Rate-region for $R_1$ and $R_2$ 
when $\{\alpha_1, \alpha_2\}  = \{1,1/4\}$ and $P_{\rm T} = 10$.
The circle mark is the solution
to \eqref{EQ:PA1} when $\bar R_1 = 2$ and $\bar R_2 = 1$.}
        \label{Fig:exam1}
\end{figure}

For example, consider the case of $K = 2$ with
$\{\alpha_1, \alpha_2\}  = \{1,1/4\}$.
If we assume that $P_{\rm T} = 10$, 
the rate-region of $R_1$ and $R_2$ can be obtained 
as in Fig.~\ref{Fig:exam1}.
We note that if $P_1 = 10$ and $P_2 = 0$, which
is the solution to \eqref{EQ:PA2}, the maximum sum rate
($\log_2 (1+ \alpha_1 P_{\rm T}) = \log_2 (11) \approx 3.459$)
is achieved.

If we consider the problem in \eqref{EQ:PA1} with
$\bar R_1 = 2$ and $\bar R_2 = 1$,
we can first decide the minimum of $P_1$ satisfying
$C_{1;1} \ge \bar R_1 = 2$, because $C_{1;1}$ is a function of only $P_1$
as in \eqref{EQ:RC1}, which is $P_1^* =  3$.
Once $P_1^*$ is decided, we can find the 
minimum of $P_2$ satisfying
$C_{2;2} \ge \bar R_2 = 1$ from \eqref{EQ:RC2}, which is $P_2^* = 7$.
We can see that $P_1^* + P_2^* = 10$. Thus, the 
solution to \eqref{EQ:PA1} can be located on the boundary
of the rate-region with $P_{\rm T} = 10$ as shown in
Fig.~\ref{Fig:exam1} (with the circle mark).

The above example demonstrates that the solution to
\eqref{EQ:PA1} can be readily found 
if the BS knows the CSI,
$\{\alpha_k\}$. In addition, the solution can achieve 
the rate-region.
Although the problem formulation in \eqref{EQ:PA1}
is attractive, the main drawback is the CSI feedback.
In \cite{Ding16}, limited feedback of CSI is considered
for a more realistic environment in NOMA.
It is also possible to perform the power allocation for NOMA with
statistical CSI. In this case, the outage probability
is usually employed as a performance measure as in
\cite{Timotheou15} \cite{Yang16} \cite{Cui16} \cite{Shi16}.

Since most power allocation methods are based on the achievable
rate, they may not be applicable when capacity achieving codes
are not employed for NOMA.
Furthermore, with fixed modulation schemes, it is not easy
to adapt the transmission rates according to the optimal powers
due to the limited flexibility.
Thus, it is often desirable to consider the power allocation
for a realistic NOMA system. To this end,
the power allocation can be investigated for a practical MUST
scheme as in \cite{Choi16}.


\section{Beamforming for Downlink NOMA}	\label{S:DB}

To increase the spectral efficiency of downlink 
in a multiuser system, 
multiuser downlink beamforming can be considered.
While there are various approaches for multiuser downlink
beamforming without NOMA, only few 
beamforming schemes are studied with NOMA. For example,
zero-forcing (ZF) approaches are considered 
in \cite{Kim13, Ding16} and
random beams are used in \cite{Higuchi13, Higuchi15}.
In \cite{Hanif16}, 
a minorization-maximization algorithm (MMA)
is employed to maximize the sum rate in NOMA with beamforming.
In \cite{Shin17}, multi-cell MIMO-NOMA networks are studied with 
coordinated beamforming.
In this section, we discuss NOMA beamforming that was studied in \cite{Kim13}.

\begin{figure}[thb]
\begin{center}
\includegraphics[width=8cm]{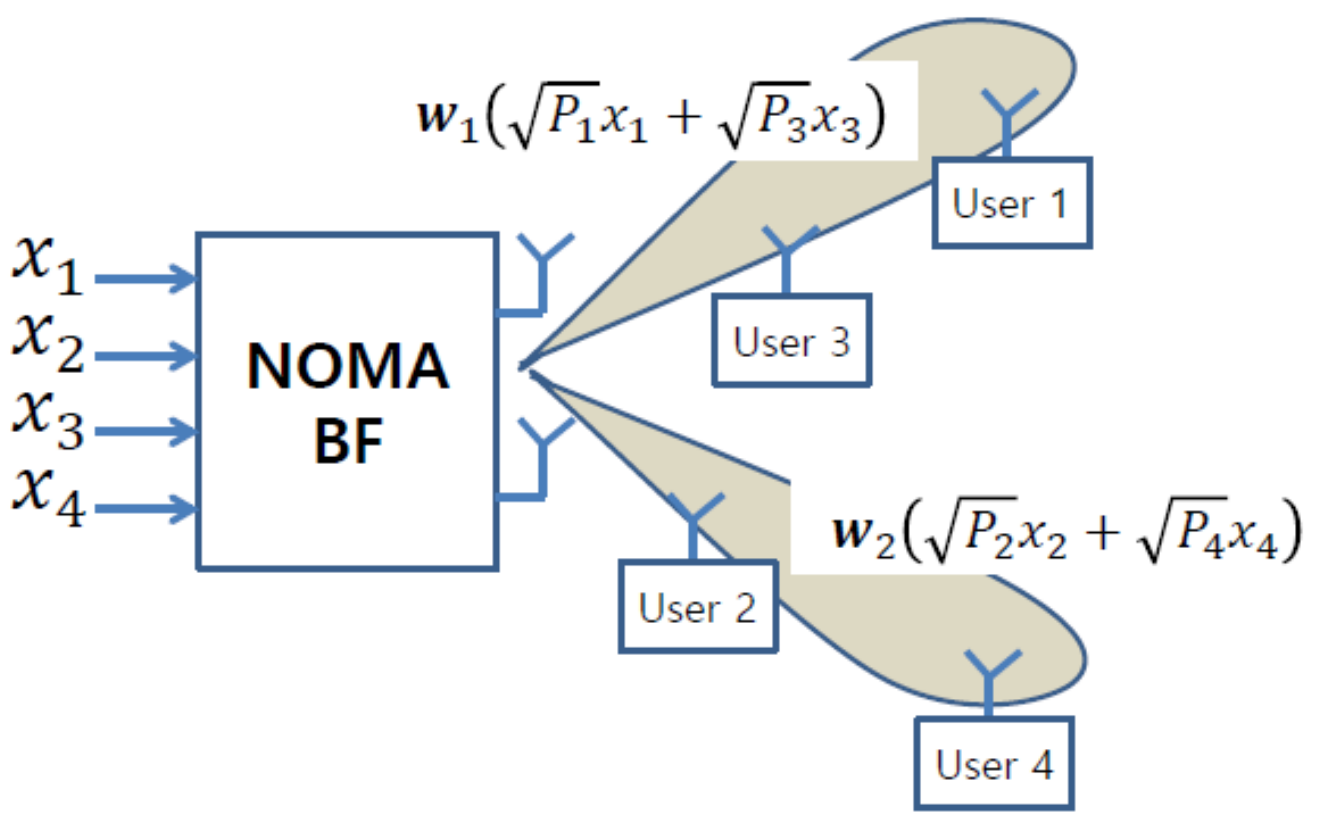}
\end{center}
\caption{An illustration of NOMA with beamforming.}
        \label{Fig:NOMA_BF}
\end{figure}

For downlink NOMA, we
can exploit the power domain
as well as the spatial domain to increase the spectral efficiency
as in \cite{Kim13}\cite{Higuchi13}.
In Fig.~\ref{Fig:NOMA_BF}, we illustrate downlink beamforming
for a system of 4 users.
There are two clusters of users. Users 1 and 3 belong to cluster 1.
In cluster 2, there are users 2 and 4. In each cluster,
the users' spatial channels should be highly correlated
so that one beam can be used to transmit signals 
to the users in the cluster \cite{Kim13}.
The resulting approach is called NOMA (downlink) beamforming.

In NOMA beamforming, there are two key problems.
The first problem is user clustering. In general, 
a group of users whose channels are highly correlated 
can form a cluster.
The second problem is beamforming.
A beam is to support a set of users in a cluster,
while this beam should not interfere with 
the users in the other clusters.
As in \cite{Kim13},
it is convenient to consider the two problems separately
at the cost of degraded performance.

To consider NOMA beamforming, we focus on one cluster
consisting of two users.
We assume that user 1 is a strong 
user (close to the BS) and user 2 is a weak user
(far away from the BS).
The signal-to-interference-plus-noise 
ratio (SINR) at user 2 is given by
\be
\gamma_2 = \frac{ |\bh_2^\rH \bw |^2 P_2}
{ |\bh_2^\rH \bw |^2 P_1 + \sigma^2}.
        \label{EQ:v1}
\ee
To keep a certain QoS, we need to satisfy
$\gamma_2 \ge G_2$,
where $G_2$ represents the target SINR at user 2
(it is assumed that if $\gamma_2 \ge G_2$, the signal to user 2
can be correctly decoded).
As usual, user 2 is assumed to be a weak user that is far away
from the BS. On the other hand, user 1,
called a strong user, is close to the BS
and able to decode the signal to user 2 and remove it by SIC,
and decode the desired signal (i.e., the signal to user 1).
Thus, at user 2, it is required that
\begin{align}
\frac{ |\bh_1^\rH \bw |^2 P_2}
{ |\bh_1^\rH \bw |^2 P_1 + \sigma^2} \ge G_2,
        \label{EQ:v3}
\end{align}
and
\begin{align}
\gamma_1 = \frac{ |\bh_1^\rH \bw |^2 P_1}
{\sigma^2} \ge G_1,
        \label{EQ:v4}
\end{align}
where $G_1$ represents the target SINR at user 1.
Note
that the sum rate of the cluster
becomes $\log_2 (1+G_1) + \log_2 (1+G_2)$.
From \eqref{EQ:v1} -- \eqref{EQ:v4},
the following set of constraints can be found:
\begin{align}
        \label{EQ:K1}
|\bh_1^\rH \bw|^2 & \ge |\bh_2 \bw|^2 \\
        \label{EQ:K2}
|\bh_1^\rH \bw|^2 P_1 & \ge G_1 \sigma^2 \\
|\bh_2^\rH \bw|^2 P_2 & \ge (
|\bh_2^\rH \bw |^2 P_1 + \sigma^2 ) G_2 .
        \label{EQ:K3}
\end{align}
Consequently, the problem
to minimize the transmit power per cluster,
$P_1 + P_2$, can be formulated with the constraints
in \eqref{EQ:K1}, \eqref{EQ:K2}, and \eqref{EQ:K3}.

\begin{figure}[thb]
\begin{center}
\includegraphics[width=\figwidth]{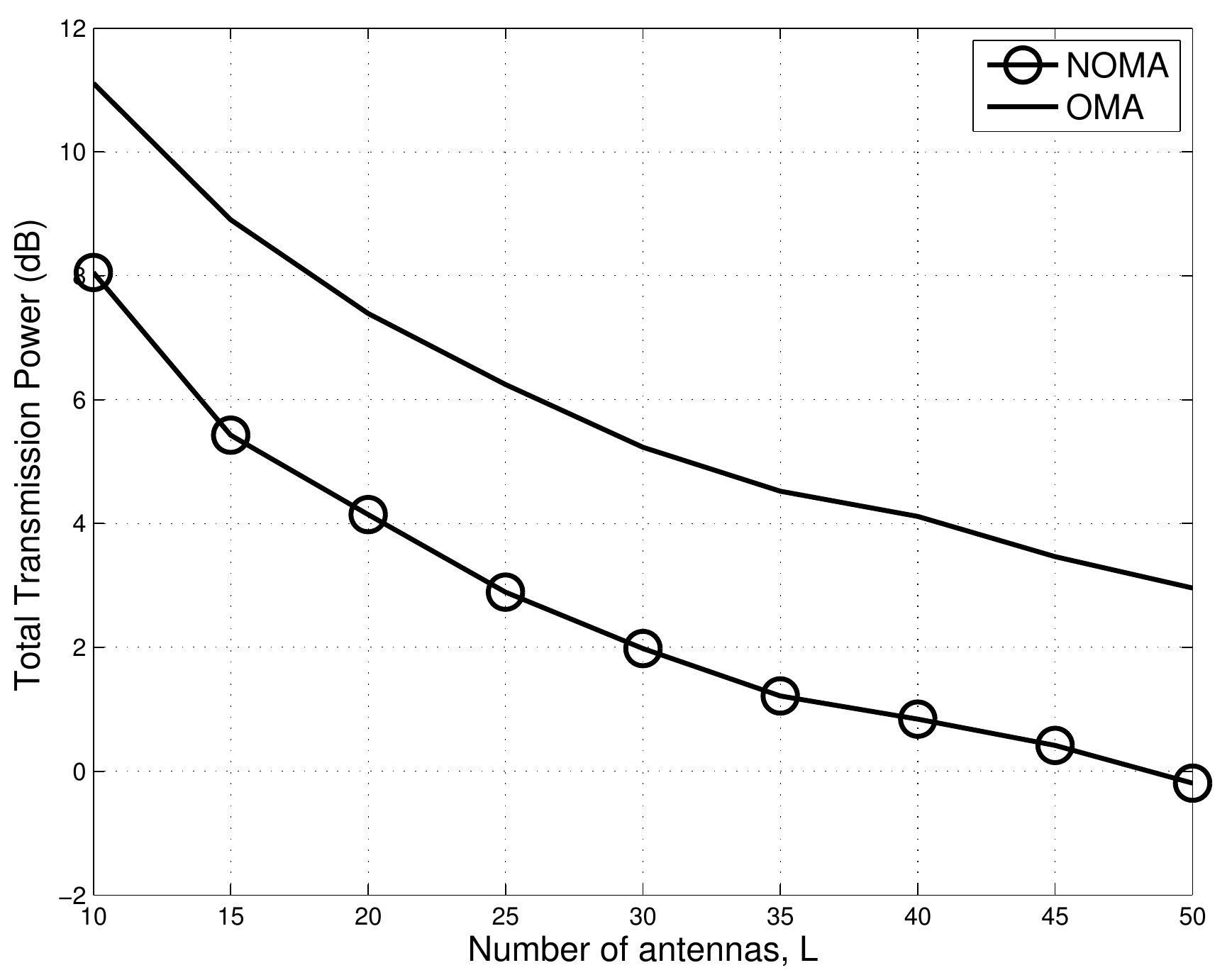}
\end{center}
\caption{The total transmit power for different values of
the number of antennas, $L$, when there
are $M = 3$ clusters with $(G_1, G_2) = (10\ {\rm dB}, 6\ {\rm dB})$.}
        \label{Fig:plt2}
\end{figure}

It is possible to find the optimal beam that minimizes
the transmission power subject to \eqref{EQ:K1} - \eqref{EQ:K3}.
With $M = 3$ clusters, the optimal beam is found 
for different numbers of antennas.
In Fig.~\ref{Fig:plt2},
we show the required total transmission power 
to meet the quality of service (QoS)
with target SINRs, $G_1$ and $G_2$.
It is clear that NOMA requires a lower transmission power
than orthogonal multiple access (OMA),
while the required total transmission power 
decreases with the number of antennas $L$
due to the beamforming gain.

In addition to NOMA beamforming, 
the user allocation or clustering plays a crucial role
in improving the performance.
In \cite{Kim13} \cite{Higuchi13}
\cite{Ding15},
other beamforming approaches
with user clustering are studied.
It is noteworthy that the NOMA beamforming approach with user clustering
is not optimal. For a better performance, 
without user clustering, as in \cite{Hanif15}, 
an optimization problem can be formulated.
However, in this case, the computational complexity
to find the optimal solution is usually high.

\section{Uplink NOMA}

NOMA can be employed for uplink transmissions
based on the coordination by the BS,
which requires signaling overhead.
It is also possible to consider uncoordinated
uplink NOMA. In this section, we briefly discuss
coordinated and uncoordinated uplink NOMA systems.

\subsection{Coordinated Uplink NOMA}

From \eqref{EQ:rt},
the mutual information between $r_t$
and $\{u_{k,t}\}$ is given by
\be
\sI (r_t; \{u_{k,t}\})
= \log_2 
\left( 1 + \sum_{k=1}^K \beta_k Q_k
\right),
	\label{EQ:Urate}
\ee
where $\beta_k= |g_k|^2$ and $Q_k = \uE[|u_{k,t}|^2]$.
In general, in order to achieve the rate
$\sI (r_t; \{u_{k,t}\})$ in 
\eqref{EQ:Urate}, the BS may need to perform joint decoding for all
$K$ signals. However, 
using the chain rule \cite{CoverBook},
it is also possible to show that
\begin{align}
\sI (r_t; \{u_{k,t}\})
& = \sum_{k=1}^K 
\sI (r_t; u_{k,t}\,|\, u_{k+1,t},\ldots u_{K,t}) \cr
& = \sum_{k=1}^K 
\log_2 
\left( 1 + \frac{\beta_k Q_k}{1+ \sum_{l = 1}^{k-1} \beta_l Q_l}
\right).
\end{align}
From this, we can clearly see that
the BS can decode $K$ signals sequentially
and independently using SIC.
For example, if $K = 2$, 
we have
\begin{align*}
\sI (r_t; \{u_{k,t}\})
= \log_2 \left( 1 + \frac{\beta_2 Q_2}{1+ \beta_1 Q_1} \right) 
+ \log_2 \left( 1 + \beta_1 Q_1 \right).
\end{align*}
Thus, if user 2 transmits the coded signals
at a rate lower than
$ \log_2 \left( 1 + \frac{\beta_2 Q_2}{1+ \beta_1 Q_1} \right) $,
the BS can decode the signals and remove them.
Then, the BS can decode the coded signals 
from user 1 at a rate lower than
$\log_2 \left( 1 + \beta_1 Q_1 \right)$.
This approach is considered to prove the 
capacity of multiple access channels \cite{CoverBook},
while CDMA and IDMA can be seen as certain implementation examples
of uplink NOMA \cite{Wang06Ping}.
In \cite{Imari14}, uplink NOMA is considered
for multicarrier systems. For a given decoding order,
the power allocation and subcarrier allocation
are carried out to maximize the sum rate.

In practice, for uplink NOMA, the BS needs to know the CSI
and decides the rates and powers according to a certain decoding order.
In other words, there could be a lot of signaling overhead
for uplink NOMA, which may offset the NOMA gain.

\subsection{Uncoordinated Uplink NOMA: Random Access with NOMA}

It is possible to employ uplink NOMA without coordination.
To this end, we can consider random access, 
e.g., ALOHA \cite{BertsekasBook}.
In ALOHA, if there are $K$ users and each of them transmits
a packet with access probability $p_a$,
the throughput becomes
$$
T = K p_a (1-p_a)^{K-1}.
$$
For a sufficiently large $K$, the throughput can be maximized
when $p_a = \frac{1}{K}$, and the maximum throughput becomes
$e^{-1} \approx 0.3679$.

\begin{figure}[thb]
\begin{center}
\includegraphics[width=\figwidth]{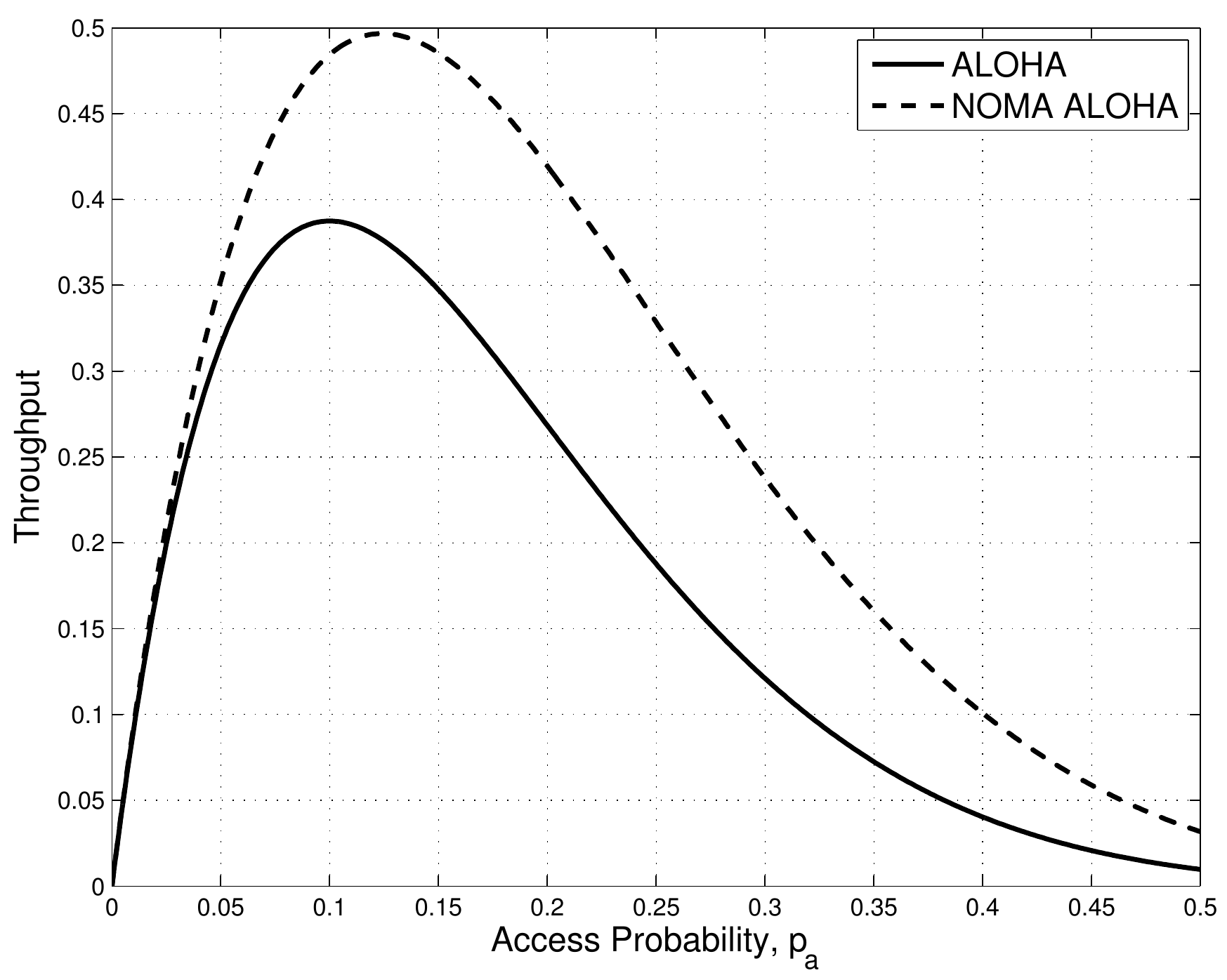}
\end{center}
\caption{Throughput of ALOHA and NOMA-ALOHA with 2 power levels
when there are $K = 10$ users.}
        \label{Fig:n_aloha}
\end{figure}

Suppose that uplink NOMA is employed and it is possible to
decode up to two users if two users have two different power levels.
Since there is no coordination, each user can choose one of two
possible power levels. In this case, the throughput becomes
$$
T = K p_a (1 - p_a)^{K-1} + 
\frac{1}{2}
\binom{K}{2} p_a^2 (1 - p_a)^{K-2},
$$
where the second term is the probability that there are two
users transmitting packets and they choose different power levels.
In Fig.~\ref{Fig:n_aloha},
we show the throughput of ALOHA and
the throughput of NOMA-ALOHA with 2 power levels
when there are $K = 10$ users.
It is clear that NOMA can improve the throughput of ALOHA.

\begin{figure}[thb]
\begin{center}
\includegraphics[width=\figwidth]{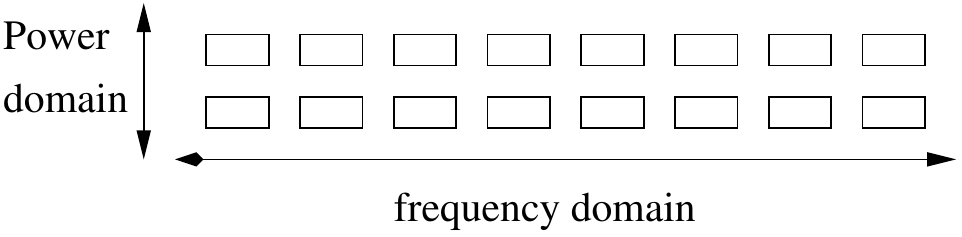}
\end{center}
\caption{Power-Frequency domain 
resource blocks for multichannel ALOHA
($L  = 2$ power levels and $B = 8$ 
(orthogonal) subcarriers.}
        \label{Fig:pdma}
\end{figure}

\begin{figure}[thb]
\begin{center}
\includegraphics[width=\figwidth]{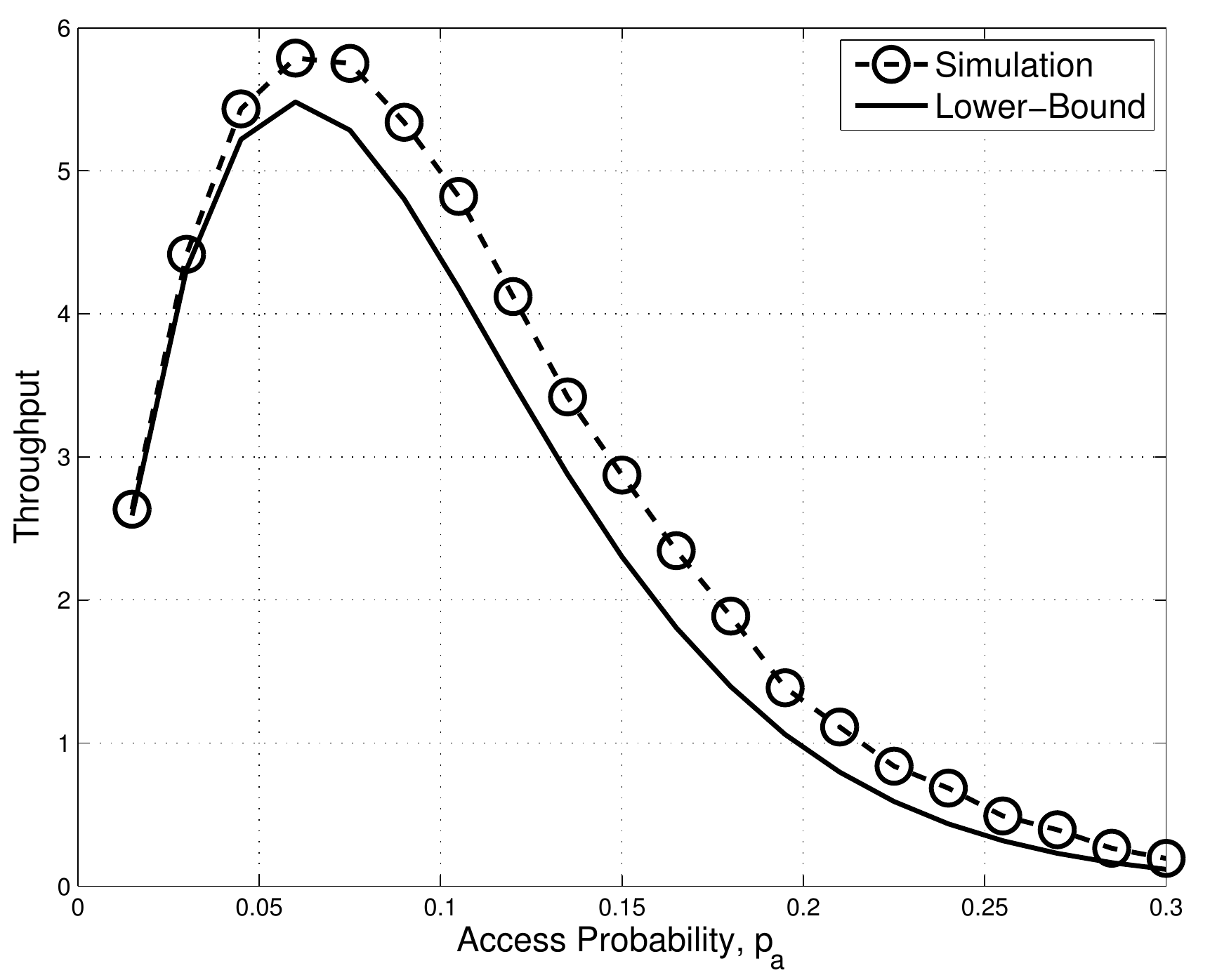}
\end{center}
\caption{Throughput for different values of access probability, 
$p_a$, when $K = 200$, $L = 4$, and $B = 6$.}
        \label{Fig:R_aplt2}
\end{figure}

We can generalize NOMA-ALOHA with
multi-channels as in \cite{ChoiJSAC}. For example,
as shown in Fig.~\ref{Fig:pdma}, the both
power and frequency domains can be considered to form multiple
sub-channels for ALOHA.
The throughput is shown 
in Fig.~\ref{Fig:R_aplt2}
when there are $K = 200$ users, $B = 6$ subcarriers,
and $L = 4$ different power levels.
It is noteworthy that the maximum throughput can be close to
$B = 6$. In other words, NOMA-ALOHA can achieve a near full utilization
of channels due to additional subchannels in the power domain.
However, the transmission power increases as a user may choose a higher power
than the required one without any MAI
\cite{ChoiJSAC}.

\section{Concluding Remarks}	\label{S:Conc}

In this paper, we presented an overview 
of NOMA as well as some recent results.
We considered the power allocation and beamforming for downlink NOMA.
We also discussed some key issues of uplink NOMA.

There are a number of topics that are not discussed in this paper,
although they are important. One of them is
user clustering. In general, user clustering is used to simplify
NOMA by applying NOMA independently to a cluster 
consisting of few users provided that inter-cluster interference
is mitigated (possibly using beamforming).
Unfortunately, the performance degradation due to user clustering
is not known, while its advantage to lower the complexity
of downlink NOMA is clear.
Another important topic to be studied is 
optimal user ordering in NOMA beamforming.
Without the space domain (or beamforming), the optimal user
ordering seems straightforward (it is usually based on the channel gains).
However, when the space 
and power domains are to be jointly exploited in NOMA,
optimal user ordering is not yet well studied.

As discussed above, although there are various issues to be addressed,
we believe that NOMA will be 
indispensable in future cellular systems.

\appendices

\section{Solution to \eqref{EQ:PA1}}	\label{A:1}

Under the assumption of \eqref{EQ:alps},
the minimum power to user 1 to satisfy $C_{1;1} \ge \bar R_1$
is given by
$$
P_1^*  = \frac{2^{\bar R_1} - 1}{\alpha_1}.
$$
While $P_1^*$ is the minimum power to guarantee the target rate for user 1,
a higher power may eventually minimize the sum power. 
However, this is not the case. To see this, let $P_1 > P_1^*$.
Then, we can show that $C_{k;k}$ with $P_1$, $k = 2,
\ldots, K$, is lower than that with $P_1^*$.
For example, we can see that
\begin{align*}
\log_2 \left(1 + 
\frac{\alpha_2 P_2}{\alpha_2 P_1 + 1}\right)
< \log_2 \left(1 + 
\frac{\alpha_2 P_2}{\alpha_2 P_1^* + 1}\right).
\end{align*}
Thus, to minimize $\sum_k P_k$, the optimal power to user 1
has to be $P_1^*$. For given $P_1^*$,
we can also find the minimum power to user 2 as follows:
$$
P_2^* = \min P_2 \ \mbox{s.t.}\ C_{2;2} \ge \bar R_2.
$$
After some manipulations, we have
$$
P_2^* = (2^{\bar R_2} - 1) \left(
P_1^* + \frac{1}{\alpha_2} \right).
$$
This is also the minimum power to user 2 that results in the minimum 
MAI to users $k > 2$. Consequently,
the minimum power for each user $k$ 
(or the solution to \eqref{EQ:PA1}) can be decided
as
\be
P_k^* = (2^{\bar R_k} - 1) \left(
\sum_{l=1}^{k-1} P_l^* + \frac{1}{\alpha_k} \right).
\ee

\bibliographystyle{ieeetr}
\bibliography{../noma}

\end{document}